\documentclass[10pt,sigconf]{acmart}

\usepackage{xpatch}

\makeatletter
\xpatchcmd{\ps@firstpagestyle}{Manuscript submitted to ACM}{}{\typeout{First patch succeeded}}{\typeout{first patch failed}}
\xpatchcmd{\ps@standardpagestyle}{Manuscript submitted to ACM}{}{\typeout{Second patch succeeded}}{\typeout{Second patch failed}}    \@ACM@manuscriptfalse
\makeatother


\usepackage{booktabs} 
 \usepackage{enumitem}
 \usepackage{balance}
\usepackage{listings}
\usepackage{placeins}

\setcopyright{none} 
\renewcommand\footnotetextcopyrightpermission[1]{} 
\begin{document}
\title{Categorization of Program Regions for Agile Compilation using Machine Learning and Hardware Support}

\author{Sanket Tavarageri}
\affiliation{%
  \institution{Intel Labs}
}
\email{sanket.tavarageri@intel.com}


\begin{abstract}

A compiler processes the code written in a high level language and produces machine executable code. The compiler writers often face the challenge of keeping the compilation times reasonable. That is because aggressive optimization passes which potentially will give rise to high performance are often expensive in terms of running time and memory footprint. Consequently the compiler designers arrive at a compromise where they either simplify the optimization algorithm which may decrease the performance of the produced code, or they will restrict the optimization to the subset of the overall input program in which case large parts of the input application will go un-optimized.

The problem we address in this paper is that of keeping the compilation times reasonable, and at the same time optimizing the input program to the fullest extent possible. Consequently, the performance of the produced code will match the performance when all the aggressive optimization passes are applied over the entire input program.

\end{abstract}

%
%

%
%

\maketitle

\section{Introduction}
We build a machine learning model to classify parts of input programs into ``easy'', and ``hard'' modules. The machine learning model will be encoded on a configurable architecture. The compiler will use the hardwired machine learning model to classify an input program into easy and hard modules and will accordingly apply optimizations – no/very little optimizations for easy ones, and aggressive optimizations for hard ones.

The differential compilation technique described here could be used to achieve dual goals: reasonable compilation times which is crucial for programmer productivity and high performance on Intel Architectures. Additionally, the machine learning model for classification of program regions and the attendant hardware support for it, will be useful for library developers, and compiler writers as well as they can readily focus on hard-to-optimize code in emerging applications and/or on emerging architectures.

\section{Overview of the Solution}

\begin{figure}[h!]
\centering
\includegraphics[scale=0.35]{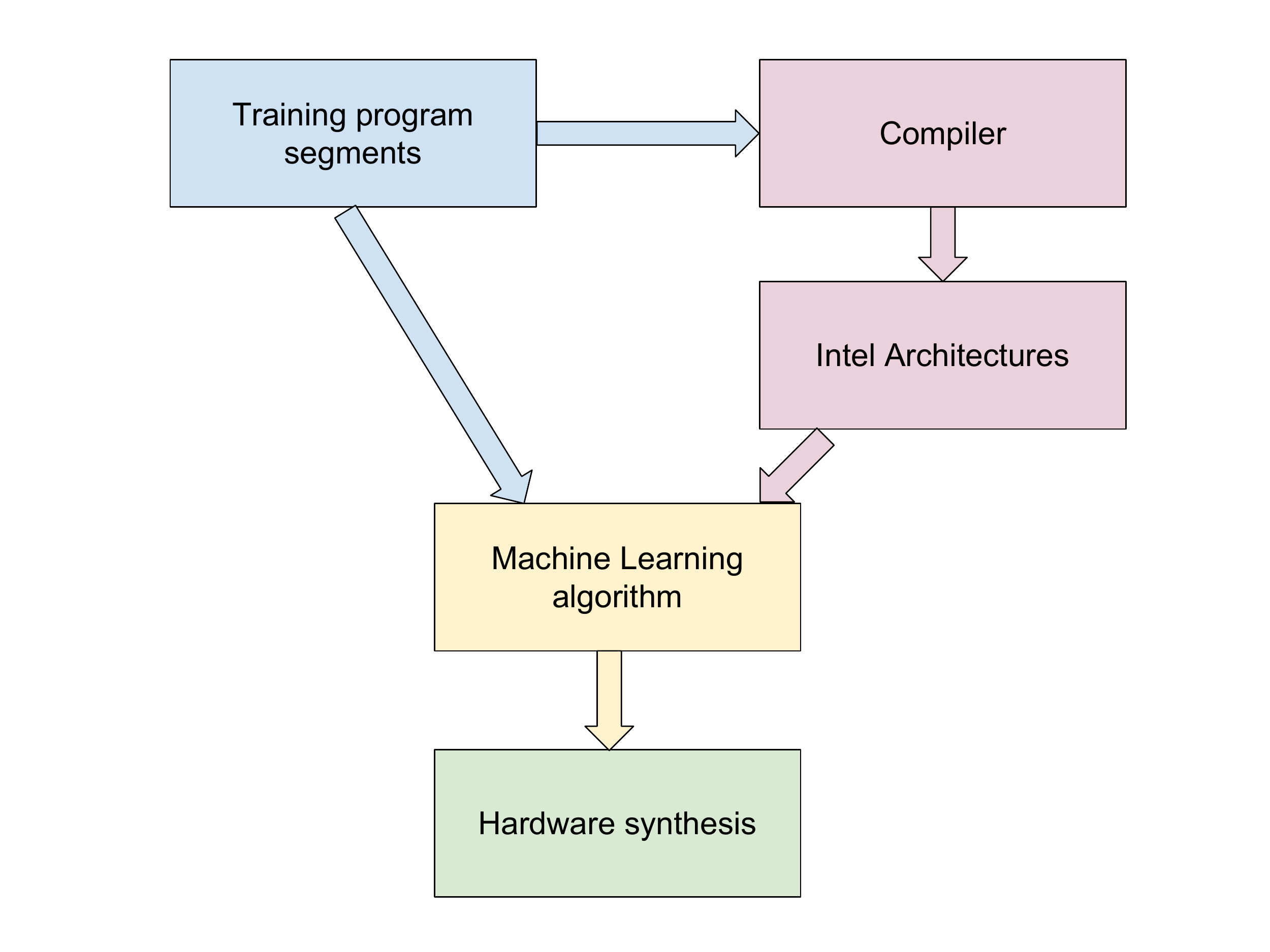}
\caption{Machine Learning training workflow}
\label{fig:ml_training}
\end{figure}

We first train a machine learning (ML) model to classify program segments into hard and easy regions. The ML model is then synthesized into hardware. Figure \ref{fig:ml_training} depicts the entire workflow. Real world code, and synthetically generated program segments are used to train the ML model. Features are extracted from programs, and form part of the input to the machine learning algorithm. The code is compiled through the compiler, and run on Intel architectures. The resulting performance statistics, and features gleaned from static analysis of code are used together to build the ML model. The model is subsequently hardwired on a configurable architecture.

\begin{figure}[h!]
\centering
\includegraphics[scale=0.35]{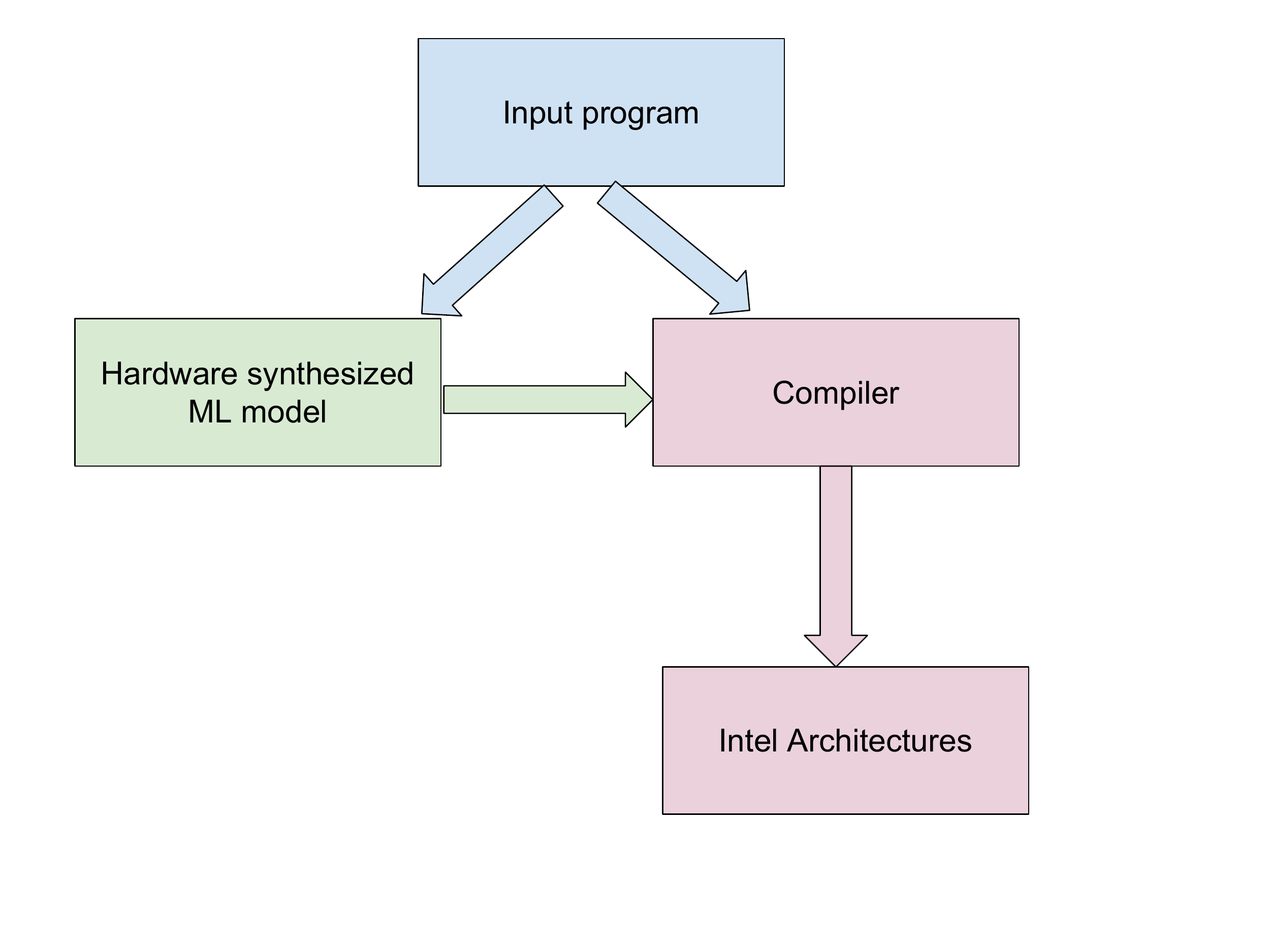}
\caption{Machine Learning model deployment}
\label{fig:ml_inference}
\end{figure}

The compiler will use the deployed ML model to perform differentiated compilation. Figure \ref{fig:ml_inference} shows the process. The input program that is to be compiled is demarcated into hard and easy modules by the trained ML model. The compiler will use this information to accelerate the compilation process by applying fewer optimizations on the easy parts, and extensive optimizations on the hard parts.

\section{Machine Learning Model for Categorization of Program Regions}
To train a machine learning model for categorization of program regions, the following tasks are performed.
1) \textbf{Collection of training data:} Both real world code and synthetically generated programs that are syntactically and semantically correct are used for training.
A system such as Csmith \cite{Yang:2011:FUB:1993498.1993532} will be used to generate synthetic code. We choose the \emph{function} as the unit of code that we will categorize as either ``easy'' or ``hard'' in terms of difficulty in optimizing it.

Each function is compiled with two optimization flags -- 1) one that represents basic optimizations (e.g., \textsf{-O1} in many compilers) 2) flag that denotes very aggressive optimizations (e.g., \textsf{-O3} in several compilers). This process creates two versions of each function: $\mathcal{F}_{basic}$ and $\mathcal{F}_{aggr}$. We execute both the versions and categorize the function as ``easy'' to optimize if the execution time of $\mathcal{F}_{basic}$ - $\mathcal{T}_{basic}$ - is very close to  that of $\mathcal{F}_{aggr}$ - $\mathcal{T}_{aggr}$. More formally, a function is labelled \textsf{easy} if the performance of $\mathcal{F}_{basic}$ is within a certain factor $\delta$ (say $80\%$) of $\mathcal{F}_{aggr}$: 

$$ \text{Label of} ~~~\mathcal{F} = \begin{cases}
\text{\textsf easy} & \text{if} ~~ \frac{\mathcal{T}_{aggr}}{\mathcal{T}_{basic}} > \delta\\
\text{\textsf hard} & \text{otherwise}
\end{cases}
$$

\textbf{2) Feature engineering:}
A static analysis of the input program is performed and the \emph{features} are extracted. It is the loops within functions that consume most execution time, and the loops need to be optimized optimally in order to achieve good performance.
We capture the characteristics of loops and the rest of the function separately to aid the machine learning algorithm in making determination on the effect of optimizations.

\begin{figure*}
\begin{lstlisting}[language=C++]
void floyd_warshall(int n, float path[N][N]) {
  int i, j, k;
  for (k = 0; k < N; k++)
    {
      for(i = 0; i < N; i++) {
        for (j = 0; j < N; j++) {
          path[i][j] = path[i][j] < path[i][k] + path[k][j] ? path[i][j] 
          : path[i][k] + path[k][j];
     }
    }
  }
}
\end{lstlisting}
\caption{The Floyd-Warshall algorithm code}
\label{fig:func}
\end{figure*}

\begin{figure*}[h!]
\centering
\includegraphics[scale=0.45]{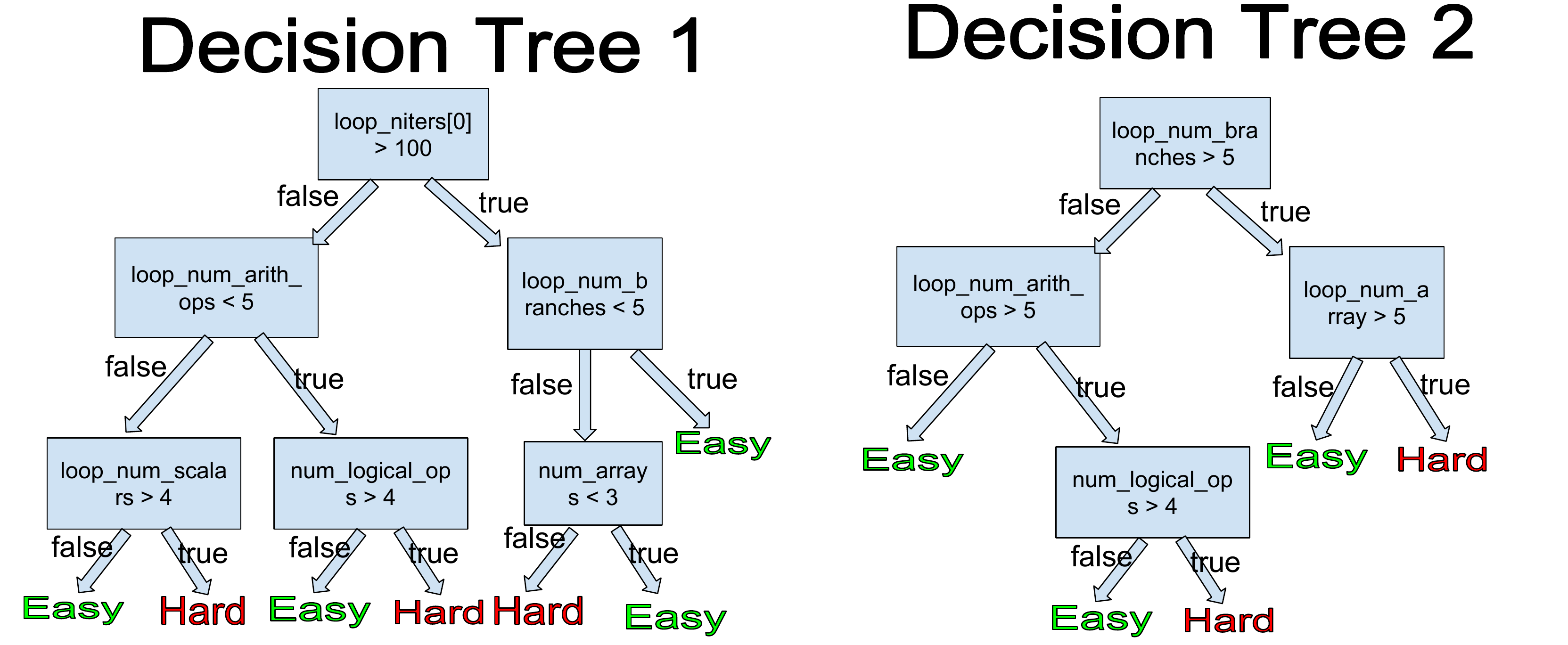}
\caption{Example random forest for classifying program regions}
\label{fig:random_forest}
\end{figure*}

We extract the following loop centric features. For each loop nest that occurs in the function, we construct the following features. When there are multiple loop nests in a function, we construct a single set of features by adding the corresponding features of different loop nests.
\begin{itemize}
\item \textit{loop\_niters}: It is a vector encoding the number of iterations of loops that appear in the function being analyzed. The tuple is ordered in terms of the textual order of loops in the program text.
\item \textit{loop\_num\_logical\_ops:} The number of logical and relational operations that appear in the loop nest.
\item \textit{loop\_num\_arith\_ops:} The definition of this is similar to that of \textit{logical\_ops} except that this feature describes the number of arithmetic operations.
\item \textit{loop\_num\_branches:} The number of branches that are present inside the loop nest.
\item \textit{loop\_num\_arrays:} The number of arrays used in the loop nest.
\item \textit{loop\_num\_scalars:} The number of scalars used in the loop nest excluding the loop variables themselves.
\end{itemize}

We \emph{reduce} the features of different loop nests using weighted addition. The weight of a loop nest is its depth -- the number of loops in the loop nest.
We \emph{normalize} the \textit{loop\_niters} feature which is a vector variable (and, it is the only vector feature, the rest are scalar features) so that we can use an element wise addition for vector addition. We determine the maximum loop depth -- \textit{max\_depth} among all loop nests within the training data and extend the \textit{loop\_niters} vectors to the \textit{max\_depth} length by padding zeros.

For non-loop code, we define the following features which have similar meaning as those defined for loop nests.

\begin{itemize}
\item \textit{num\_logical\_ops}
\item \textit{num\_arith\_ops} 
\item \textit{num\_branches} 
\item \textit{num\_arrays}
\item \textit{num\_scalars}
\end{itemize}

\textbf{Example:}
Consider the function shown in Fig. \ref{fig:func} which forms the core of Floyd–Warshall algorithm for finding the shortest paths between all pairs of vertices in a graph.

The features developed for the above function will be as follows.
\begin{itemize}
\item \textit{loop\_niters}: [S, S, S]. \\
If the loop iterations are symbolic constants as in this case, they are mapped to the special attribute `S'. 
\item \textit{loop\_num\_logical\_ops:} 1. \\
The comparator operator `<' is the only relational operator that appears in the loop nest (excluding the operators used in the \textsf{for} loops). 
\item \textit{loop\_num\_arith\_ops:} 1. \\
 `+' is the only arithmetic operation that is executed in any branch of the inner most loop.
\item \textit{loop\_num\_branches:} 1. \\
This corresponds to the \textsf{ternary if} operator in the loop.
\item \textit{loop\_num\_arrays:} 1. \\
The array -- \textsf{path} is the array used.
\item \textit{loop\_num\_scalars:} 0.
\end{itemize}

The values of all non-loop features  are 0 as there is no non-loop code in the \textsf{floyd\_warshall} function.

\textbf{3) ML model selection and training:}
We use the Random forests \cite{breiman2001random} as the machine learning model for classification of program regions. A random forest comprises of multiple decision trees, each of which will predict whether a given function will be ``easy'' or ``hard'' to optimize. The majority vote of the trees is considered the overall prediction of the random forest. The random forest is superior to single trees as it avoids the problem of overfitting the data -- each constituent tree is trained with the subset of the training data, and consequently the random forest avoids overfitting.

An example random forest comprising of two decision trees is shown in Figure \ref{fig:random_forest}. The nodes in a decision tree are conditionals that are functions on the features. The features of the function to be classified are extracted and are then used to run them through the component decision trees. Each tree will label the function as either \emph{easy} or \emph{hard} and the majority vote is considered the true label of the function.

\FloatBarrier

\textbf{4) Discussion:}
We described an exemplar approach to realizing the creation of a machine learning model to categorize program regions. We note that there are a great many variations possible in terms of selection of features, and selection of the machine learning model itself. For example, from a feature selection point of view, one could use contextual flow graphs \cite{ben2018neural} that combine data- and control- flow graphs of programs. With regards to selection of machine learning models, deep neural networks could be used. But the approach described in this document represents a viable implementation of the idea of using machine learning models for the purpose of classifying program regions.

\section{Related Work}

In the JIT compilation approach \cite{JIT_wiki}, a program is interpreted. However “hot” loops (loops where significant running time is spent, or the ones that are frequently executed) are identified at runtime, and those loops are compiled to native code. Virtual Machines such as Java Virtual Machine (JVM) use variants of this strategy to achieve high performance for interpreted languages.

Leather et al \cite{leather2009automatic} develop a machine learning approach to program optimization, specifically for choosing loop unroll factors. Many other research efforts have also focused on using different machine learning methods to optimize programs in various ways. Wang et al's article \cite{wang2018machine} provides a comprehensive survey of such attempts.

The previous solutions (such as JIT compilation) attempt to identify hot loops at runtime, and compile them. This approach can accelerate compilation process by compiling only those loops that are important. However, the overall achieved performance is not optimal because 1) the hot loops are identified only when the execution is well underway, and not apriori 2) the non-hot loops are not compiled at all, but are interpreted.

The machine learning based compiler optimizations focus on improving the performance of the generated code, but in fact increase the compilation time significantly. Since the compilation can take inordinately long, the solutions are often not practical, and are rarely used in practice.

\section{Conclusion}
In this paper, we proposed a machine learning based solution to accelerate the process of compilation of programs. Compiler writers often face the challenge of designing sophisticated compiler passes to increase the performance of generated code while keeping the compilation times reasonable. In this paper, the machine learning approach to categorization of program regions as easy or hard addresses that problem by automatically figuring out which program segments benefit from longer compiler passes, and which do not. The upshot of the differentiated compiler treatment of program regions will be that expensive compiler passes are applied only to the regions that very likely benefit from them and skipping the passes for regions that do not. Consequently, the performance of the generated code will be high and the compilation times will be practical.

\balance

\bibliographystyle{ACM-Reference-Format}
\bibliography{paper}

\end{document}